\documentclass[article,amsmath,amssymb]{revtex4-2}
\usepackage{graphicx}
\usepackage{dcolumn}
\usepackage[colorlinks]{hyperref}
\usepackage{bm}

\begin{document}
\title{Subradiant Decay in 2D and 3D Atomic Arrays}

\author{Nicola Piovella}
\affiliation{Dipartimento di Fisica "Aldo Pontremoli", Universit\`{a} degli Studi di Milano, Via Celoria 16, I-20133 Milano, Italy \&
INFN Sezione di Milano, Via Celoria 16, I-20133 Milano, Italy}
\author{Romain Bachelard}
\affiliation{Departamento de Física, Universidade Federal de São Carlos, Rodovia Washington Luís, km 235 - SP-310, 13565-905 São Carlos, SP, Brazil}

\begin{abstract}
Subradiance is a phenomenon where coupled emitters radiate light at a slower rate than independent ones. While its observation was first reported in disordered cold atom clouds, ordered subwavelength arrays of emitters have emerged as promising platforms to design highly cooperative optical properties based on dipolar interactions. In this work we characterize the eigenmodes of 2D and 3D regular arrays, using a method which can be used for both infinite and very large systems. In particular, we show how finite-size effects impact the lifetimes of these large arrays. Our results may have interesting applications for quantum memories and topological effects in ordered atomic arrays.
\end{abstract}

\maketitle

\section{Introduction}

Dicke superradiance corresponds to the enhancement of the radiation rate by $N$ emitters due to coherence effects~\citep{Dicke1954,Gross1982}, as opposed to subradiance for which it is inhibited~\citep{Pavolini1985,Crubellier1987}. Recently, the subradiant suppression of radiative decay has spurred strong interest due to its potential applications for quantum memories~\citep{Facchinetti2016,Jen2016}, excitation transfer~\citep{Needham2019,Cech2023} and topological photonics~\citep{Bettles2017}. Many-atom subradiance has been observed almost ten years ago in disordered clouds of cold atoms, with long lifetimes evidencing the phenomenon~\cite{Guerin2016}.

More recently, ordered sub-wavelength arrays of atoms have emerged as a promising configuration to achieve subradiance, with distinct advantages over disordered ensembles for applications such as photon storage or photonic gates~\citep{Porras2008,Jenkins2012,Bettles2016a,Shahmoon2017,Asenjo2017,Jenkins2017}. In these systems, the emitters are regularly arranged with a spatial period below the atomic transition wavelength, so the dipolar interactions lead to strong cooperative optical properties. 

In this context, infinite lattices offer interesting insights on the scattering properties of these arrays. The single-excitation eigenmodes of the scattering problem are translationally invariant and obey Bloch's theorem. This allows one to define the quasi-momentum $\mathbf{k}$ and calculate explicitly the collective frequency shift and decay rate of each of these modes~\cite{Asenjo2017}. Differently, in finite systems these collective features can only be accessed by numerical diagonalization, leading to severe limitations on the system size which can be simulated. Furthermore, most studies on subradiance in ordered systems have focused on one-dimensional (1D) atomic chains~\cite{Jen2016,Needham2019,Cech2023,Bettles2016b,Das2020,Ferioli2021,Zoubi2010,Bettles2015} and two-dimensional (2D) arrays~\cite{Bettles2017,Shahmoon2017,Asenjo2017,Rui2020,Bettles2016a}, with three-dimensional configurations less investigated ~\cite{Brechtelsbauer2021,Sierra2022,Ruks2025}. We note that while the band structure of 2D and 3D arrays have been studied through the density of states~\cite{Coevorden1996,Antezza2009,Perczel2017}, the connection between this phenomenon and subradiance has not been yet clarified.

In this work, we investigate the cooperative decay rates of 2D and 3D atomic array, using a technique where the discrete sums over the emitters is turned into an angular integral. It provides the cooperative rate for the quasi-modes in infinite arrays~\cite{Piovella2024}. But it is also particular convenient to study the rates of large size systems, for which the technique provides a good approximation, and numerical diagonalization are out of reach. Used successfully for 1D systems~\cite{Piovella2024}, we show the collective decay rates of the associated generalized Dicke state (also called generalized Bloch state~\cite{Zhang2020}) can also be derived explicitly for 2D and 3D arrays. The work thus opens new perspectives for the study of cooperative scattering in large, regular atomic systems.

\section{Microscopic modeling of atomic arrays}

\subsection{Coupled Dipole Model}

Let us consider $N$ atoms with transition frequency $\omega_0=ck_0$ between the ground state $|g_j\rangle$ and the excited state $|e_j\rangle$ ($j=1,\dots,N$), linewidth $\Gamma_0=k_0^3d_{ge}^2/(3\pi\epsilon_0\hbar)$, where $\hbar$ is the Planck constant, $\epsilon_0$ the vacuum permittivity and $d_{ge}$ the electric-dipole transition matrix element. The atoms are at fixed positions $\mathbf{r}_j$ and have transition dipole moments ${\textbf{d}_j}$, with $\mathbf{d}_j=d_{ge}\mathbf{\hat d}_j$. In the Born-Markov approximation, one can trace out the quantized light fields and obtain the field-mediated dipole-dipole couplings between the atoms described by an effective Hamiltonian \cite{Akkermans2008}:
\begin{equation}
H=-\hbar\sum_{j,m=1}^N \mathbf{\hat d}^*_j\cdot \mathbf{G}(\mathbf{r}_j-\mathbf{r}_m)\cdot \mathbf{\hat d}_m\sigma^\dagger_j\sigma_m,\label{Heff}
\end{equation}
where $\sigma_j=|g_j\rangle\langle e_j|$ is the lowering operator for atom $j$. The dyadic Green's tensor $\mathbf{G}$ for the electromagnetic field in the 3D free space in vacuum is given by
\begin{equation}
\mathbf{G}(\mathbf{r})=\frac{3\Gamma_0}{2}\frac{e^{ik_0r}}{k_0r}\left\{\mathbb{I}-\mathbf{\hat n}\mathbf{\hat n}+\left(\mathbb{I}-3\mathbf{\hat n}\mathbf{\hat n}\right)\left(\frac{i}{k_0r}-\frac{1}{k_0^2r^2}\right)
\right\},\label{dyadic}
\end{equation}
where $\mathbb{I}$ is the identity tensor $3\times 3$, $r=|\mathbf{r}|$ and $\mathbf{\hat n}=\mathbf{r}/r$. The dyadic Green's tensor (\ref{dyadic}) can be obtained from the scalar Green's function $G(r)=\exp(ik_0r)/(k_0r)$ \cite{JOSA2014}
\begin{equation}
\mathbf{G}(\mathbf{r})=\frac{3\Gamma_0}{2}\left[\mathbb{I}+\frac{1}{k_0^2}\nabla\nabla\right]G(r).
\label{dyadic:2}
\end{equation}
If we assume that all transition dipoles point in the same direction, $\mathbf{\hat d}_j=\mathbf{\hat d}$, such as in presence of a strong external field, the interaction term between atoms $j$ and $m$ read
\begin{equation}
G_{jm}=\mathbf{\hat d}^*_j\cdot \mathbf{G}(\mathbf{r}_{jm})\cdot \mathbf{\hat d}_m=
\frac{3\Gamma}{2}\left[1+\frac{1}{k_0^2}(\mathbf{\hat d}\cdot\nabla_{\mathbf{r}_{jm}})^2\right]\frac{e^{ik_0r_{jm}}}{k_0r_{jm}},
\end{equation}
with $\mathbf{r}_{jm}=\mathbf{r}_j-\mathbf{r}_m$ and $r_{jm}=|\mathbf{r}_{jm}|$.
Note that $G_{jm}$ has both a real and imaginary part, which describe the photon exchange between the atoms and their collective emission out of the system, respectively. Throughout this work we focus our attention on the imaginary part of $G_{jm}$, defining
\begin{eqnarray}
\Gamma_{jm}&=&\mathrm{Im}[G_{jm}]=
\frac{3\Gamma_0}{2}\left[1+\frac{1}{k_0^2}(\mathbf{\hat d}\cdot\nabla_{\mathbf{r}_{jm}})^2\right]\frac{\sin(k_0r_{jm})}{k_0r_{jm}}.
\label{Gamma:jm}
\end{eqnarray}
Introducing the angular average
\begin{eqnarray}
\left\langle f(\theta_0,\phi_0)\right\rangle_\Omega=\frac{1}{4\pi}\int_0^{2\pi}d\phi_0\int_0^\pi \sin\theta_0 f(\theta_0,\phi_0)d\theta_0,\nonumber
\end{eqnarray}
and using the property $\langle e^{-ix\cos\theta}\rangle_\Omega =(\sin x)/x$, we can rewrite $\Gamma_{jm}$ as an angular average of the photon exchange term between the two atoms
\begin{eqnarray}
\Gamma_{jm}&=&
\frac{3\Gamma_0}{2}\left\langle
\left[1-(\mathbf{\hat d}\cdot\mathbf{\hat k}_0)^2\right]
e^{-i\mathbf{k}_0\cdot(\mathbf{r}_{j}-\mathbf{r}_{m})}
\right\rangle_\Omega
\label{Gamma},
\end{eqnarray}
where $\mathbf{k}_0=k_0(\sin\theta_0\cos\phi_0,\sin\theta_0\sin\phi_0,\cos\theta_0)$ and $\mathbf{\hat k}_0=\mathbf{k}_0/k_0$. 
The remarkable property of Eq.~\eqref{Gamma} is that the contribution of the two atoms can now be factorized, which will allow us to calculate explicitly the collective decay rates, as we will later see.

\subsection{Generalized Dicke state.}

We hereafter consider regular atomic arrays with lattice step $d$, with atoms at positions $\mathbf{r}_{j_x,j_y,j_z}=d[(j_x-1)\mathbf{\hat e}_x+(j_y-1) \mathbf{\hat e}_y+(j_z-1)\mathbf{\hat e}_z]$ with $j_x=1,\dots N_x$, $j_y=1,\dots,N_y$ and $j_z=1,\dots,N_z$, so $N=N_xN_yN_z$ is the total number of atoms. We then introduce the generalized Dicke states~\cite{Piovella2024}, or generalized Bloch states \cite{Zhang2020}:
\begin{equation}
|\mathbf{k}\rangle =\frac{1}{\sqrt{N}}\sum_{j=1}^N e^{i\mathbf{k}\cdot \mathbf{r}_j}|j\rangle.\label{Dicke}
\end{equation}
The single-excitation states are $|j\rangle=|g_1,\dots, e_j,\dots, g_N\rangle$ and $\mathbf{k}\in {\cal S}$, where ${\cal S}$ is the volume of first Brillouin zone in the reciprocal lattice, i.e., $|k_x|,|k_y|,|k_z|\le \pi/d$. It includes the symmetric Dicke state~\cite{Dicke1954} for $\mathbf{k}=0$~, as well as  the timed-Dicke state introduced in Ref.~\cite{Scully2006,Scully2015} for a driving with wavevector $\mathbf{k}=\mathbf{k}_0$. The set $\{|\mathbf{k}\rangle\}$ satisfies the completeness relation
\begin{equation}
\left(\frac{d}{2\pi}\right)^3\int_{{\cal S}}d^3k |\mathbf{k}\rangle\langle \mathbf{k}|=1.
\end{equation}
For a finite lattice these states are not orthogonal since
\begin{equation}
\langle \mathbf{k}'|\mathbf{k}\rangle=\frac{\sin[(k_x-k_x')dN_x/2]}{\sin[(k_x-k_x')d/2]}
\frac{\sin[(k_y-k_y')dN_y/2]}{\sin[(k_y-k_y')d/2]}\frac{\sin[(k_z-k_z')dN_z/2]}{\sin[(k_z-k_z')d/2]}e^{i(\mathbf{k}-\mathbf{k}')\cdot \mathbf{r}_{N_x,N_y,N_z}}.
\end{equation}
nevertheless, in the limit of an infinite lattice the asymptotic property $\langle \mathbf{k}'|\mathbf{k}\rangle\rightarrow \delta^{(3)}(\mathbf{k}-\mathbf{k}')$ turns the set $\{|\mathbf{k}\rangle\}$ into an orthonormal basis for the single-excitation manifold.

\subsection{Collective decay rate}

The collective decay rate of mode $\mathbf{k}$ is defined as~\cite{Piovella2024}
\begin{equation}
\Gamma(\mathbf{k})=-\frac{1}{\hbar}\mathrm{Im}\langle \mathbf{k}|H|\mathbf{k}\rangle=
\frac{1}{N}\sum_{j=1}^N\sum_{m=1}^N\Gamma_{jm}e^{i\mathbf{k}\cdot \mathbf{r}_{jm}}.\label{FT}
\end{equation}
Eq.~(\ref{FT}) corresponds to the Fourier transform of $\Gamma_{jm}$ only if the lattice is infinite, when modes~(\ref{Dicke}) are the Bloch states and set $\{|\mathbf{k}\rangle\}$ is discrete. Nevertheless, treating $\mathbf{k}$ as a continuous variable in Eq.~(\ref{FT}) is a good approximation for the mode when $N$ is sufficiently large~\cite{Asenjo2017}. We hereafter refer to $\Gamma(\mathbf{k})$ as the ``continuous spectrum'' of the decay rate. We highlight that we are not referring to the discrete spectrum of the eigenvalues of the system, described by the non-Hermitian Hamiltonian~\eqref{Heff}, but rather to the imaginary part of the expectation value of the effective Hamiltonian on the collective state of Eq.~\eqref{Dicke}, describing the interference among the (single-photon) fields emitted by $N$ atoms. In particular, in the infinite $N$ limit, $\Gamma(\mathbf{k})$ is precisely the decay rate of mode $\mathbf{k}$, yet in the finite-$N$ case, the exact rate must be obtained by diagonalizing the full matrix~\eqref{Heff}.

Inserting the cross decay term \eqref{Gamma} into \eqref{FT} leads to
\begin{eqnarray}
\Gamma(\mathbf{k})&=&
\frac{3\Gamma_0}{2N}\sum_{j=1}^N\sum_{m=1}^N
\left\langle
\left[1-(\mathbf{\hat d}\cdot\mathbf{\hat k}_0)^2\right]
e^{i(\mathbf{k}-\mathbf{k}_0)\cdot \mathbf{r}_{jm}}
\right\rangle_\Omega\nonumber\\
&=&\frac{3\Gamma_0}{2N}
\left\langle
\left[1-(\mathbf{\hat d}\cdot\mathbf{\hat k}_0)^2\right]
|F(\mathbf{k})|^2
\right\rangle_\Omega,\label{Gamma:k}
\end{eqnarray}
where we have introduced the structure factor
\begin{equation}
F(\mathbf{k})=\sum_{j=1}^N e^{i(\mathbf{k}-\mathbf{k}_0)\cdot \mathbf{r}_j}.\label{F}
\end{equation}
Using this explicit equation, let us now investigate the decay rates for a 2D square lattice and a 3D cubic lattice.

\section{Two-dimensional square lattice}

\subsection{General expression of the decay rate}

For a 2D atom array in the plane $(x,y)$, with $\mathbf{r}_j=d[(j_x-1)\mathbf{\hat e}_x+(j_y-1)\mathbf{\hat e}_y]$ with $j_x=1,\dots N_x$ and $j_y=1,\dots,N_y$, the rate~\eqref{Gamma:k} reads
\begin{eqnarray}
  \Gamma(k_x,k_y) &=&\frac{3\Gamma_0}{2N_xN_y}
  \left\langle
\left[1-(\mathbf{\hat d}\cdot\mathbf{\hat k}_0)^2\right]
|F(k_x,k_y)|^2
\right\rangle_\Omega,\label{Gamma:2D}
\end{eqnarray}
where the structure factor now factorizes for each direction:
\begin{equation}
F(k_x,k_y)=\left(\sum_{j_x=1}^{N_x} e^{i(k_x-k_0\sin\theta_0\cos\phi_0)d(j_x-1)}\right)
\left(\sum_{j_y=1}^{N_y} e^{i(k_y-k_0\sin\theta_0\sin\phi_0)d(j_y-1)}\right).\label{Fxy}
\end{equation}
 Summing over $j_x$ and $j_y$, we obtain
\begin{equation}
|F(k_x,k_y)|^2=\frac{\sin^2[(k_x-k_0\sin\theta_0\cos\phi_0)dN_x/2]}{\sin^2[(k_x-k_0\sin\theta_0\cos\phi_0)d/2]}\times
\frac{\sin^2[(k_y-k_0\sin\theta_0\sin\phi_0)dN_y/2]}{\sin^2[(k_y-k_0\sin\theta_0\sin\phi_0)d/2]}.
\end{equation}
The property
\begin{equation}
\frac{\sin^2(Nt)}{\sin^2t}\approx N^2\sum_{m=-\infty}^{+\infty} \mathrm{sinc}^2\left[\left(t-m\pi\right)N\right]\label{sinc}
\end{equation}
can then be used, along with the assumption of large $N_x$ and $N_y$, to write the decay rate as
\begin{eqnarray}
  \Gamma(k_x,k_y) &=&\frac{3\Gamma_0}{8\pi}N_xN_y\sum_{(m_x,m_y)\in\mathbb{Z}^2} 
  \int_0^{2\pi}d\phi_0\int_0^\pi d\theta_0\sin\theta_0\nonumber\\
  &\times&
  \left[
  1-(\hat d_x\sin\theta_0\cos\phi_0+\hat d_y\sin\theta_0\sin\phi_0+\hat d_z\cos\theta_0)^2
  \right]\nonumber\\
  &\times&\mathrm{sinc}^2\left[\left(k_x-g_x-k_0\sin\theta_0\cos\phi_0\right)\frac{dN_x}{2}\right]\nonumber\\
  &\times&\mathrm{sinc}^2\left[\left(k_y-g_y-k_0\sin\theta_0\sin\phi_0\right)\frac{dN_y}{2}\right]\label{Gamma:2D}.
\end{eqnarray}
where $g_\alpha=2\pi m_\alpha/d$, $\alpha=x,\ y$, are the components of the reciprocal lattice vector $\mathbf{g}$. Note that different values of $\mathbf{g}$ correspond to different Brillouin zones. In order to perform the integration over the angles $\theta_0$ and $\phi_0$, we introduce the variables
$v_x=(k_x-g_x-k_0\sin\theta_0\cos\phi_0)/2$ and $v_y=(k_y-g_y-k_0\sin\theta_0\sin\phi_0)/2$, which leads to
\begin{eqnarray}
\Gamma(k_x,k_y)&=&\frac{3\Gamma_0}{\pi k_0^2}N_xN_y
\sum_{m_x,m_y}\int dv_x\int dv_y \frac{\mathrm{sinc}^2(v_xdN_x)\mathrm{sinc}^2(v_ydN_y)}{\sqrt{1-C^2(v_x,v_y)}}\nonumber\\
&\times&
\left\{1-\left[\hat d_xC_x(v_x)+\hat d_yC_y(v_y)+\hat d_z\sqrt{1-C^2(v_x,v_y)}\right]^2\right\}.
\label{Gamma:2D:1}
\end{eqnarray}
We have here defined
\begin{eqnarray}
C_x(v_x)&=&(k_x-g_x-2v_x)/k_0,\\
C_y(v_y)&=&(k_y-g_y-2v_y)/k_0,
\end{eqnarray}
so that
$C^2(v_x,v_y)=C_x^2(v_x)+C_y^2(v_y)<1$. The final expression forthe rate then depends on the lattice size, $(N_x,N_y)$, and we first discuss the case of an infinite 2D system.

\subsection{Infinite square array}

\begin{figure}
     \centerline{\scalebox{0.5}{\includegraphics{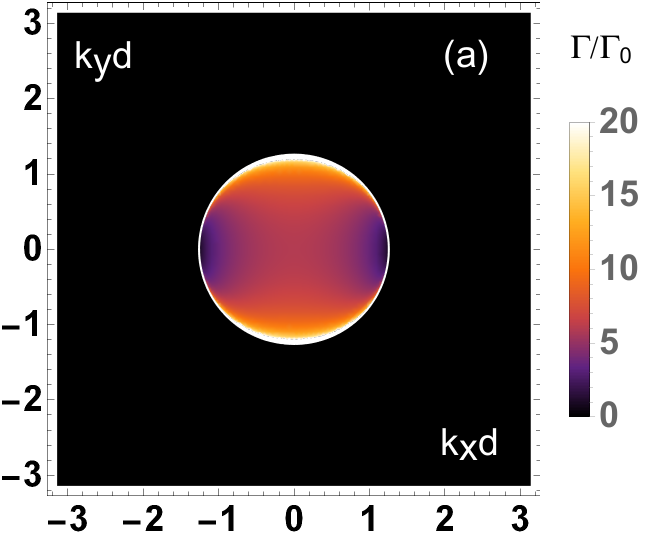}\includegraphics{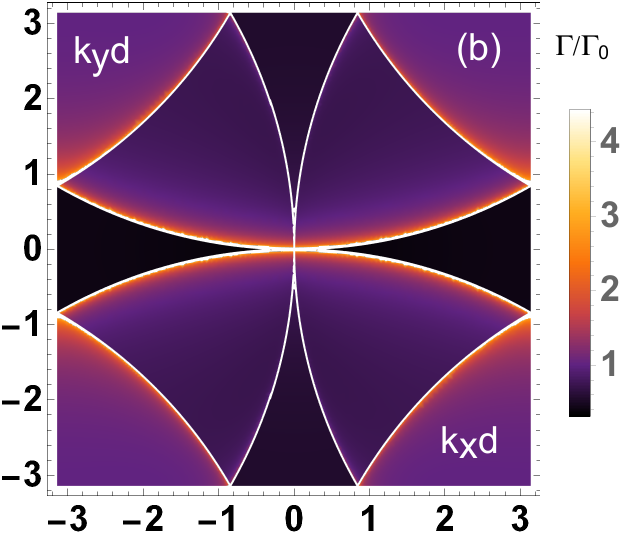}}}
        \caption{Collective decay rate $\Gamma/\Gamma_0$ in the $\mathbf{k}=(k_xd, k_yd)$ plane for the infinite array with (a) $d=\lambda_0/5$ and (b) $d=\lambda_0$, and dipoles oriented along the $x$-axis.}
        \label{2Dplot1}
\end{figure}

In the limit $N_x,N_y\rightarrow\infty$, using the property $\mathrm{sinc}^2(v_{x,y}dN_{x,y})\rightarrow (\pi/dN_{x,y})\delta(v_{x,y})$ in the rate expression~(\ref{Gamma:2D:1}) leads to 
\begin{eqnarray}
\Gamma(k_x,k_y)&=&\frac{3\Gamma_0\pi}{(k_0d)^2}
\sum_{m_x,m_y}
\frac{1}{\sqrt{1-C^2(0,0)}}\left(1-\left[\hat d_xC_x(0)+\hat d_yC_y(0)+\hat d_z\sqrt{1-C^2(0,0)}\right]^2\right),\nonumber\\
\label{Gamma:2D:infinite}
\end{eqnarray}
with the condition $C^2(0,0)<1$, i.e., $(k_x-g_x)^2+(k_y-g_y)^2<k_0^2$: Each term of the sums in Eq.~(\ref{Gamma:2D:infinite}) is different from zero inside a circle of radius $k_0$ and center in $\mathbf{g}$. This allows us to recover the result of Ref.~\cite{Asenjo2017} for parallel polarization, with the dipoles oriented in the array plane ($\hat d_z=0$):
\begin{eqnarray}
\Gamma^\parallel(\mathbf{k})&=&\frac{3\Gamma_0\pi}{(k_0d)^2}
\sum_{\mathbf{g}}
\frac{k_0^2-[\mathbf{\hat d}\cdot(\mathbf{k}-\mathbf{g})]^2}{\sqrt{k_0^2-|\mathbf{k}-\mathbf{g}|^2}},
\label{Gamma:2D:infinite:paral}
\end{eqnarray}
as well as for perpendicular polarization, with the dipoles orthogonal to the 2D array plane ($\hat d_x=\hat d_y=0$):
\begin{eqnarray}
\Gamma^\perp(\mathbf{k})&=&\frac{3\Gamma_0\pi}{(k_0d)^2}
\sum_{\mathbf{g}}
\frac{|\mathbf{k}-\mathbf{g}|^2}{\sqrt{k_0^2-|\mathbf{k}-\mathbf{g}|^2}},
\label{Gamma:2D:infinite:perp}
\end{eqnarray}
with $\mathbf{g}=(g_x,g_y)$ the reciprocal array vector introduced previously. Fig.~\ref{2Dplot1} shows $\Gamma(k_x,k_y)/\Gamma_0$ as a function of $k_xd$ and $k_yd$ for $d=\lambda_0/5$ and $d=\lambda_0$ ($\lambda_0=2\pi/k_0$ is the atomic transition wavelength), and for dipoles oriented along the $x$-axis. The black areas correspond to dark states, with $\Gamma(k_x,k_y)=0$: They have a suppressed emission, akin to subradiant modes, yet reaching $\Gamma(k_x,k_y)=0$ in this infinite-size limit. Note that for $d<\lambda_0/2$ only the terms $g_x=0$ and $g_y=0$ contribute to the sums in Eq.~\eqref{Gamma:2D:infinite}, while for $\lambda_0/2<d<\lambda_0$, the terms $g_x=\pm 2\pi/d$ and $g_y=\pm 2\pi/d$ also contribute.

\subsection{Finite square array}

\begin{figure}
     \centerline{\scalebox{0.5}{\includegraphics{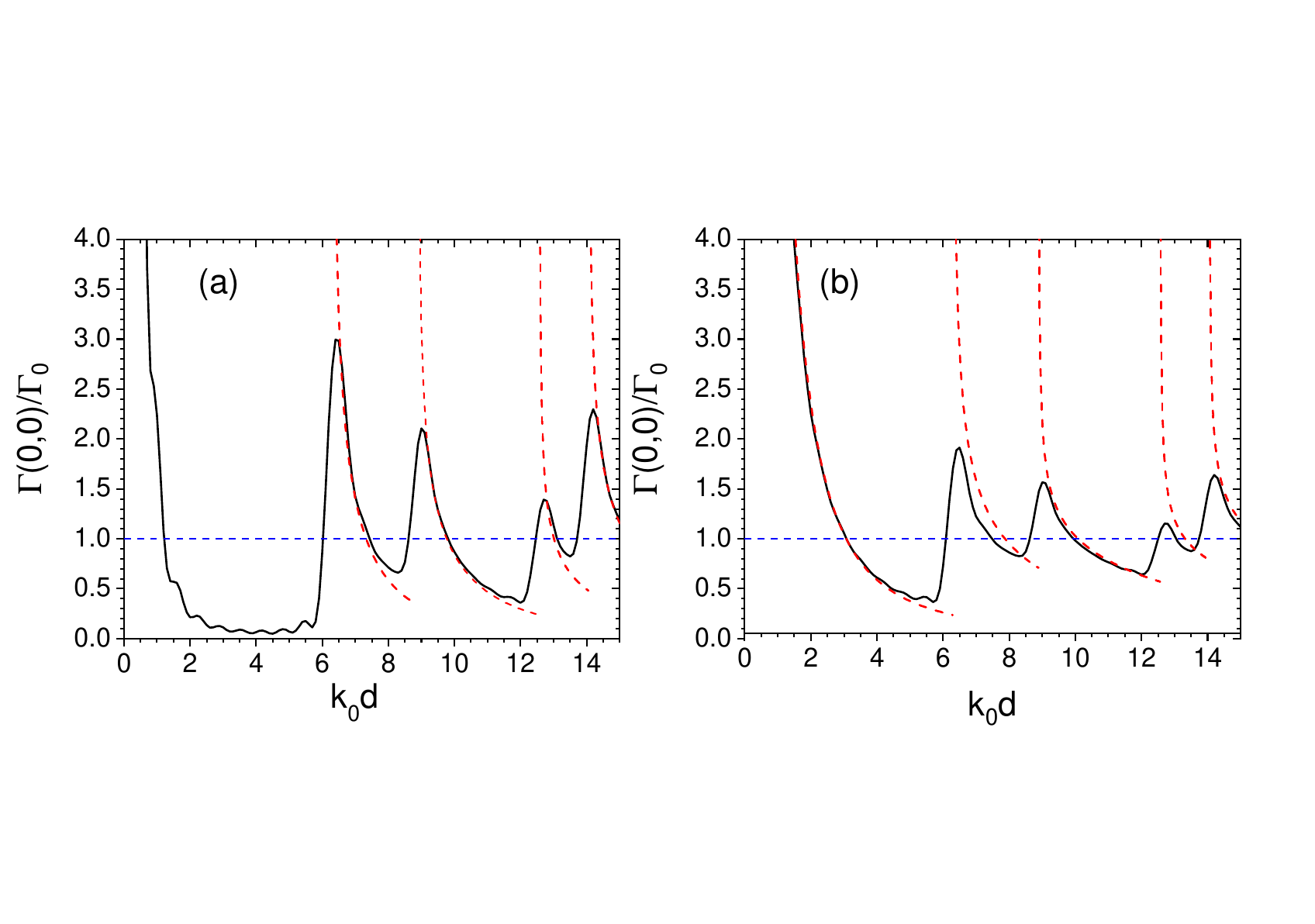} }}
        \caption{$\Gamma(0,0)/\Gamma_0$ vs $k_0d$ for a square $N\times N$ lattice with $N=10$ and polarization (a) perpendicular to the plane, $\mathbf{\hat d}=(0,0,1)$, and (b) within the atomic plane, $\mathbf{\hat d}=(1,0,0)$. The dashed red lines are the solution for an infinite lattice, provided by Eq.~\ref{Gamma:2D:infinite}). In the case (a), $\Gamma(0,0)=0$ when $0<k_0d<\pi$ for an infinite lattice.}
        \label{fig2}
\end{figure}

The collective decay rates for a finite square are obtained from integral~(\ref{Gamma:2D}), for finite $N_x$ and $N_y$. A particular case is that of the mode $(k_x,k_y)=(0,0$), which can be addressed, for example, using a laser with wave vector perpendicular to the atomic plane, as done in Ref.~\cite{Rui2020}. 
Fig.~\ref{fig2} shows $\Gamma(0,0)/\Gamma_0$ as a function of the normalized step $k_0d$, for a  square array of $N\times N=10\times 10$ atoms, with dipole polarization $\mathbf{\hat d}=(0,0,1)$ for panel (a) and $\mathbf{\hat d}=(1,0,0)$ for panel (b). 

The dashed red lines are the solution for an infinite lattice, see Eq.~(\ref{Gamma:2D:infinite}), and we observe that they provide a good estimate of the decay rates, apart from the strongly superradiant or subradiant ones. In particular, the rate of the symmetric mode goes to $\Gamma(0,0)=N^2\Gamma_0$ for a vanishing step $k_0d\rightarrow 0$ (out of scale in the figure): This corresponds to the superradiant case in the subwavelength regime, which scales as the number of particles~\cite{Dicke1954}.

As for subradiance, for perpendicular polarization [panel (a) in Fig.~\ref{fig2}], the collective decay rate is close to zero for subwavelength array step, $0<k_0d<2\pi$ (i.e., $d<\lambda_0$). Differently, for parallel polarization [panel (b)] the decay rate is less than the single-atom decay one $\Gamma_0$ only for $\pi<k_0d<2\pi$, and its value is overall much larger. The origin of this much more efficient suppression of the emission in the perpendicular polarization channel can be found in the pattern of the dipole radiation, which is in the atomic plane in that case so the light remains confined in the system, whereas for parallel polarization the dipole radiation has a substantial component out of the atomic array, thus promoting stronger leaks of the radiation out of the lattice. 

\subsection{Large-$N$ limit for the finite square array}

For a large but finite lattice ($1\ll N_x,\ N_y < \infty$), and considering modes along the $x$ direction ($k_y=0$) and with dipoles perpendicular to the 2D array ($\mathbf{\hat d}=\hat{z}$), an approximate expression for the subradiant region of the spectrum $\Gamma(k_x,k_y)$ can be derived. Indeed, assuming $k_x>k_0$ and $d<\lambda_0/2$ in the integral~(\ref{Gamma:2D:1}) we obtain, for $\mathbf{g}=0$ and up to terms $O(1/N^2_{x,y})$ (see Appendix~\ref{Appendix:A}):
\begin{eqnarray}
\Gamma(k_x,0)&=&\frac{3\pi\Gamma_0}{2(k_0d)^3\sqrt{k_xd}}\left\{k_xd\sqrt{N_x}
\frac{\sin\left[\frac{1}{2}\arctan\left(\frac{1}{v_0}\right)\right]}{(1+v_0^2)^{1/4}}
-\frac{4\sqrt{2}}{\sqrt{N_x}}
\sqrt{\frac{v_0+\sqrt{1+v_0^2}}{1+v_0^2}}\right\},\nonumber\\
\label{Gamma:2D:approx}
\end{eqnarray}
where $v_0=(dN_x/4k_x)(k_x^2-k_0^2)$. In the limit $dN_x\gg 4k_x/(k_x^2-k_0)^2$ (that is, far from the subradiant boundary $|k_x|=k_0$), Eq.~(\ref{Gamma:2D:approx}) can be approximated by
\begin{eqnarray}
\Gamma(k_x,0)&\approx &\frac{6\pi\Gamma_0}{(k_0d)^3N_x}
\frac{k_x(2k_0^2-k_x^2)}{(k_x^2-k_0^2)^{3/2}},
\label{Gamma:2D:approx:2}
\end{eqnarray}
which is valid for $k_0<|k_x|<\sqrt{2}k_0$. Hence, in the subradiant region ($|k_x|>k_0$) the rate scales as $\Gamma(k_x,0)\sim 1/N_x$. 
In the superradiant region $|k_x|<k_0$, for large $N_x$ the expression is close to the solution for an infinite chain~(\ref{Gamma:2D:infinite}), which diverges for $|k_x|=k_0$. Indeed, for $|k_x|=k_0$ and large $N_x$, Eq.~(\ref{Gamma:2D:approx}) approximates as $\Gamma(k_0,0)\approx 3\pi\Gamma_0 \sqrt{N_x/(2k_0d)^3}$.
 
\begin{figure}
      \centerline{\scalebox{0.3}{\includegraphics{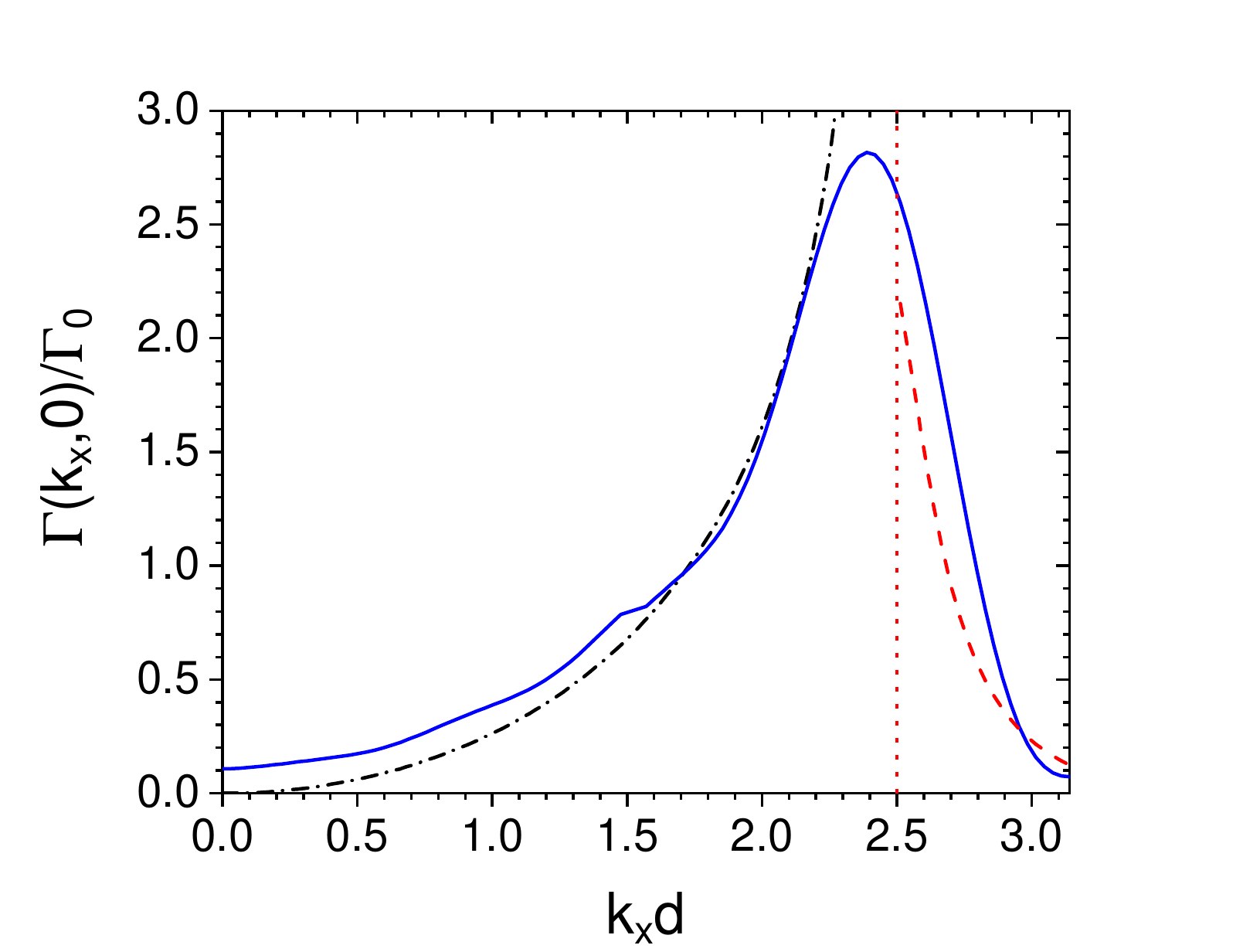}}}
        \caption{Collective decay rate $\Gamma(k_x,0)/\Gamma_0$ as a function of $k_xd$, for a square array with $d=0.8\lambda_0$ and $N_x=10$, with a polarization orthogonal to the array ($\mathbf{\hat d}=(0,0,1)$). Full blue line: exact solution~(\ref{Gamma:2D:1}); Red dashed line: approximate solution~(\ref{Gamma:2D:approx}); Dash-dotted line: solution for the infinite chain~(\ref{Gamma:2D:infinite}). The vertical dotted line stands for the value $k_x=k_0$.}
        \label{2Dplot3}
\end{figure}

These features are illustrated in Fig.~\ref{2Dplot3}, where the exact solution of $\Gamma(k_x,0)/\Gamma_0$ from Eq.~(\ref{Gamma:2D}) [continuous blue line] is compared with the approximate solution~(\ref{Gamma:2D:approx}) [dashed red line] as a function of $k_xd$, in the region $0<k_x<\pi/d$. The dash-dotted black line is the infinite chain solution~(\ref{Gamma:2D:infinite}), with the vertical dotted line marking the value $k_x=k_0$. 

\subsection{Finite square array: radial mode distribution}

For an infinite square array with perpendicularly oriented dipoles [$\mathbf{\hat d}=(0,0,1)$], the rate $\Gamma(k_x,k_y)$ presents a radial symmetry around $(m_x,m_y)=(0,0)$  -- see Eq~.(\ref{Gamma:2D:infinite:perp}). Let us exploit this symmetry, now for a finite array, with $N_x=N_y=N\gg 1$. Substituting $k_x=k_\perp\cos\theta$, $k_y=k_\perp\sin\theta$, $v_xdN=v\cos\theta'$ and $v_ydN=v\sin\theta'$ into Eq.~(\ref{Gamma:2D:1}), we then use the approximation $\mathrm{sinc}^2(v_{x}dN)\mathrm{sinc}^2(v_{y}dN)\approx 1/(1+v^4/4)$. For a subwavelength step $d<\lambda_0/2$, it leads to
\begin{eqnarray}
\Gamma(k_\perp)&=&\frac{3\Gamma_0}{\pi (k_0d)^2}
\int_{0}^\infty\frac{v}{1+v^4/4}\int_0^{2\pi}\frac{C^2(v,\theta)}{\sqrt{1-C^2(v,\theta)}}d\theta dv,
\label{Gamma:r}
\end{eqnarray}
where $C(v,\theta)=\sqrt{(k_\perp/k_0)^2-4(k_\perp/k_0)(v/k_0dN)\cos\theta+(2v/k_0dN)^2}$.

\begin{figure}
            \centerline{\scalebox{0.5}{\includegraphics{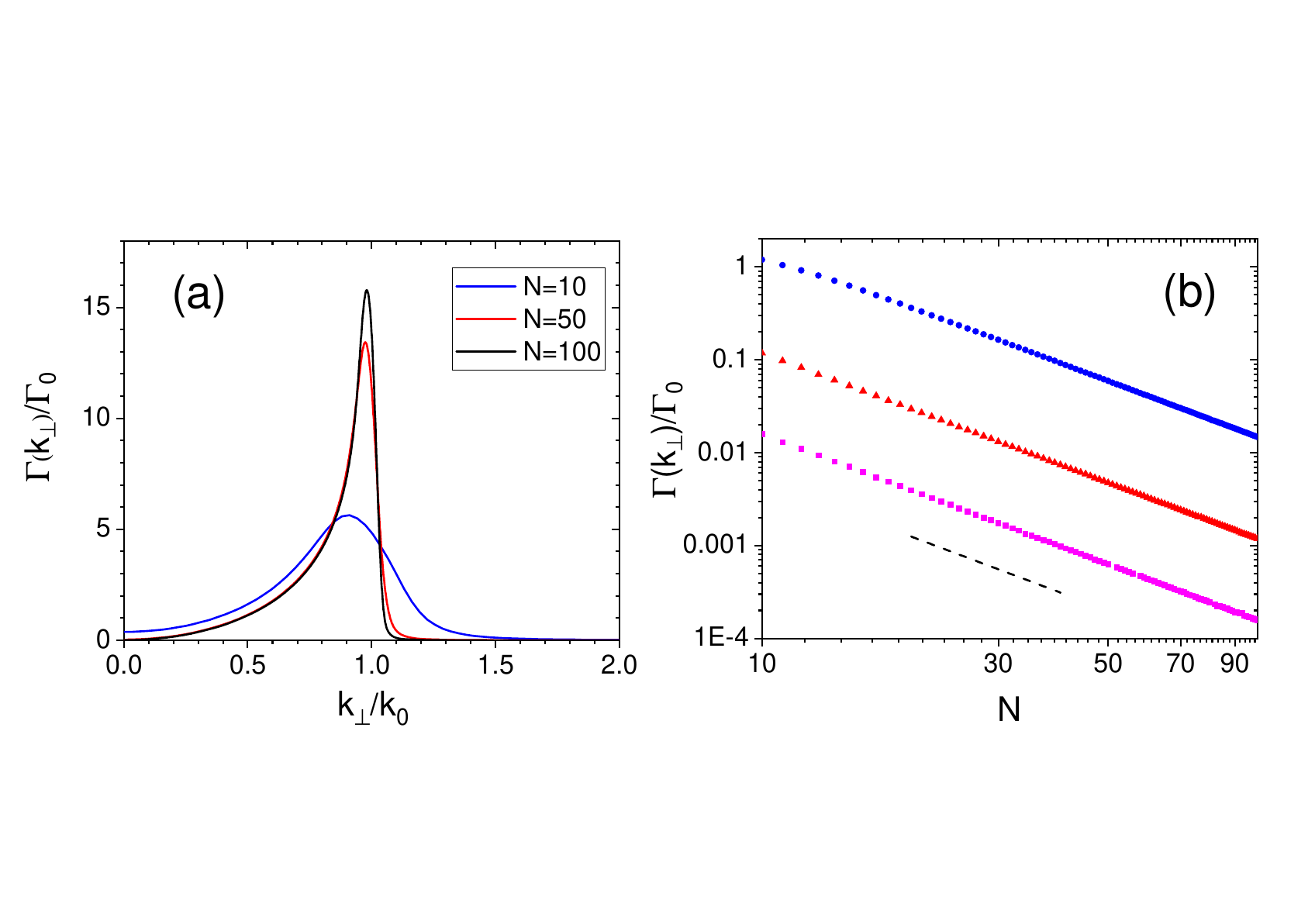}}}
        \caption{Collective decay rate $\Gamma(k_\perp)/\Gamma_0$ as a function (a) of $k_\perp/k_0$ for a square array  and $N=10$ (blue line), $N=50$ (red line) and $N=100$ (black line), and (b) of $N$ for $k_\perp/k_0=1.2$ (pink boxes), $1.5$ (red triangles) and $2$ (black circles); the dashed line in (b) shows the $1/N^{2}$ dependence. All simulations realized for $d=\lambda_0/4$.}
        \label{2Dplot5}
\end{figure}

Fig.\ref{2Dplot5}(a) shows the behavior of $\Gamma(k_\perp)$ for $d=\lambda_0/4$ and different values of $N$, from a numerical integration of Eq.~(\ref{Gamma:r}). 
If we now introduce $v'=v/N$ in Eq.~(\ref{Gamma:r}), it now takes the form
\begin{eqnarray}
\Gamma(k_\perp)&=&\frac{3\Gamma_0 N^2}{\pi (k_0d)^2}
\int_{0}^\infty\frac{v'}{1+N^4v'^4/4}\int_0^{2\pi}\frac{C^2(v',\theta)}{\sqrt{1-C^2(v',\theta)}}d\theta dv'\label{Gamma:r:2}
\end{eqnarray}
where $C(v',\theta)=\sqrt{(k_\perp/k_0)^2-4(k_\perp/k_0)(v'/k_0d)\cos\theta+(2v'/k_0d)^2}$ is independent on $N$. This allows us to deduce that in the large $N$ limit, Eq.~(\ref{Gamma:r:2}) scales as $1/N^2$. This analysis is confirmed by the numerical analysis of Eq.~(\ref{Gamma:r}), shown in Fig.\ref{2Dplot5}(b), where the $1/N^2$ trend is observed already for $N\geq 10$. 

\section{3D cubic atom array}

Let us now consider a 3D cubic array, with atoms at positions $\mathbf{r}_j=d[(j_x-1)\mathbf{\hat e}_x+(j_y-1)\mathbf{\hat e}_y+(j_z-1)\mathbf{\hat e}_z]$ with $j_x=1,\dots N_x$, $j_y=1,\dots,N_y$ and $j_z=1,\dots,N_z$. The rate of mode $\mathbf{k}(k_x,k_y,k_z)$~(\ref{Gamma:k}) now writes
\begin{eqnarray}
  \Gamma(\mathbf{k}) &=&\frac{3\Gamma_0}{2N_xN_yN_z}
  \left\langle
\left[1-(\mathbf{\hat d}\cdot\mathbf{\hat k}_0)^2\right]
|F(\mathbf{k})|^2
\right\rangle_\Omega\label{Gamma:3D},
\end{eqnarray}
where  
\begin{eqnarray}
|F(\mathbf{k})|^2&=&\frac{\sin^2[(k_x-k_0\sin\theta_0\cos\phi_0)dN_x/2]}{\sin^2[(k_x-k_0\sin\theta_0\cos\phi_0)d/2]}\\
&\times &\frac{\sin^2[(k_y-k_0\sin\theta_0\sin\phi_0)dN_y/2]}{\sin^2[(k_y-k_0\sin\theta_0\sin\phi_0)d/2]}\times\frac{\sin^2[(k_z-k_0\cos\theta_0)dN_z/2]}{\sin^2[(k_z-k_0\cos\theta_0)d/2]}.\nonumber
\end{eqnarray}
Assuming a large array, $N_x,N_y,N_z\gg 1$, and using the same transformation as in Eq.~(\ref{Gamma:2D:1}) for 2D arrays, the rate rewrites
\begin{eqnarray}
\Gamma(\mathbf{k})&=&\frac{3\Gamma_0}{2\pi k_0^2}N_xN_yN_z
\sum_{\mathbf{g}}\int dv_x\int dv_y\mathrm{sinc}^2(v_xdN_x)\mathrm{sinc}^2(v_ydN_y)\nonumber\\
&\times &
\frac{1-\left[\hat d_xC_x(v_x)+\hat d_yC_y(v_y)+\hat d_z\sqrt{1-C^2(v_x,v_y)}\right]^2}{1-\sqrt{1-C^2(v_x,v_y)}}\nonumber\\
&\times &\left\{\mathrm{sinc}^2\left[\left(k_z-g_z-k_0\sqrt{1-C^2(v_x,v_y)}\right)\frac{dN_z}{2}\right]\right.\nonumber\\
&+&\left.\mathrm{sinc}^2\left[\left(k_z-g_z+k_0\sqrt{1-C^2(v_x,v_y)}\right)\frac{dN_z}{2}\right]\right\}
\label{Gamma:3D:3}
\end{eqnarray}
where  $C^2(v_x,v_y)<1$. Here, $\mathbf{g}=(2\pi/d)\mathbf{m}$ is the reciprocal lattice vector, with $\mathbf{m}=(m_x, m_y, m_z)\in\mathbb{Z}^3$.

\subsection{Infinite 3D array}

In the infinite array limit, $N_x, N_y, N_z\rightarrow\infty$,   one can use the property $\mathrm{sinc}^2(v_{\alpha}dN_{\alpha})\rightarrow (\pi/dN_{\alpha})\delta(v_{\alpha})$ to simplify rate~(\ref{Gamma:3D:3}) as
\begin{eqnarray}
\Gamma(\mathbf{k})&=&\frac{3\pi^2\Gamma_0}{2\pi k_0^2d^3}
\sum_{\mathbf{g}}
\left\{1-\left[\mathbf{\hat d}\cdot\left(\mathbf{k}-\mathbf{g}\right)\right]^2\right\}
\delta\left[\left|\mathbf{k}-\mathbf{g}\right|-k_0\right].
\label{Gamma:3D:infinite}
\end{eqnarray}
Each term of the sum in Eq.~(\ref{Gamma:3D:infinite}) is thus an infinitely-thin shell of center $\mathbf{g}$ and radius $k=k_0$.
 
We note that in Ref.~\cite{Asenjo2017} subradiant modes in infinite 1D and 2D arrays have been described as "guided", since the associated radiation fields are evanescent in the directions transverse to the array. In contrast, in 3D arrays the radiation modes are extended over space, in all directions. Thus, differently from the 1D and 2D cases, the subradiant modes in finite 3D arrays cannot be identified using the results of the infinite arrays, as it has been possible in 1D and 2D arrays. This has lead to focusing on the numerical diagonalization of the scattering matrix to determine the eigenvalues in the finite-$N$ case~\cite{Asenjo2017}. In our approach, subradiant modes in a finite 3D array have a finite width around the surface $k=k_0$, as we will discuss in the next section.

\subsection{Decay rate for mode along an axis of the finite cubic array}

\begin{figure}
      \centerline{\scalebox{0.4}{\includegraphics{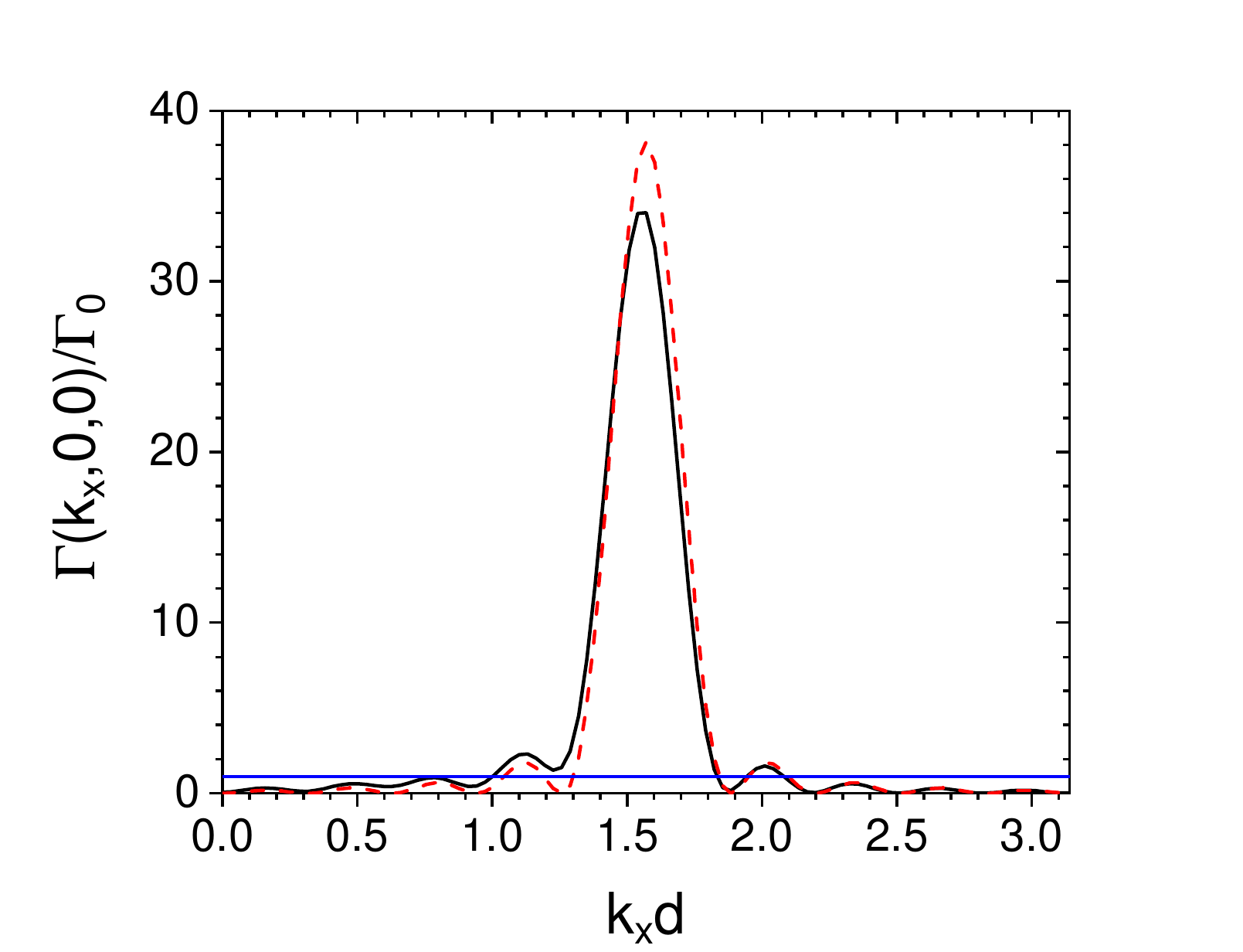}}}
        \caption{Collective decay rate $\Gamma(k_x,0,0)/\Gamma_0$ as a function of $k_xd$ for a cubic array with step $d=\lambda_0/4$ and size $N_x=N_y=N_z=20$, as calculated from the exact expression~(\ref{Gamma:3D}) (full black line) and from the approximated expression~(\ref{Gamma:3D:approx}) (red dashed line).}
        \label{3D}
\end{figure}

Considering a cubic array with a large size, $N_x, N_y, N_z\gg1$, an approximate solution for the decay rate $\Gamma(k_x,k_y,k_z)$ can be derived from Eq.~(\ref{Gamma:3D:3}). In the subradiant region of the spectrum, $k>k_0$ with $\mathbf{g}=0$ and considering for instance modes with polarization along the $z$ axis and along the $k_x$ one ($\mathbf{k}=(k_x,0,0)$), we derive, up to terms $O(1/N^2_{x,y,z})$ (see Appendix \ref{Appendix:B}):
\begin{eqnarray}
\Gamma(k_x,0,0)&=&\frac{3\pi\Gamma_0 N_x}{2(k_0d)^2}\mathrm{sinc}^2\left(\frac{dN_x(k_x-k_0)}{2}\right).
\label{Gamma:3D:approx}
\end{eqnarray}
Note that for $k_x=k_0$, using that $N=N_xN_yN_z$ is the total number of atoms in the cubic array with edges $L_y=N_yd$ and $L_z=dN_z$, the rate rewrites $\Gamma(k_x,0,0)=2b_0\Gamma_0$ where $b_0=(3\pi/4) N/(k_0^2L_yL_z)$ is the resonant optical thickness. In contrast, $\Gamma(k_x,0,0)$ decreases as $1/N_x$ when $k_x$ moves away from $k_0$, both inside and outside the shell, toward the dark zone. In Fig.~\ref{3D} the rate $\Gamma(k_x,0,0)/\Gamma_0$ as a function of $k_xd$ shows the very good agreement between the exact expression~(\ref{Gamma:3D}) and the approximate one~(\ref{Gamma:3D:approx}).

In conclusion, for an infinite 3D array subradiance appears almost everywhere, except when $|\mathbf{k}-\mathbf{g}|=k_0$. For a finite cubic array, subradiant states are close to those of the infinite array, but with the sharp surfaces of the spectrum turned into smooth ones, with tails in the dark regions whose widths decrease as the inverse of the atom number $N$.

\section{Discussion}

We have thus characterized the decay rates of 2D and 3D atomic arrays, identifying families of Bloch states with strong subradiance properties. In particular, we have shown that using an integral representation to deal with the sum over the many emitters allows one to obtain explicit expressions for the rates, but also to study finite-size effects in different dimensions. With the recent realization of a subradiant 2D atomic array mirror~\citep{Rui2020}, the development of techniques to study systems too large to be studied directly by numerical diagonalization is a crucial task.

Another important aspect is the preparation of the generalized Dicke state $|\mathbf{k}\rangle$ of Eq.(\ref{Dicke}). For example, in Ref.~\citep{Rui2020} the state $\mathbf{k}=0$ has been excited by a laser perpendicular to a 2D atomic array with $k_0d\approx4.27$, such that weak subradiance has been observed (see Fig.\ref{fig2}b). More generally, a possible strategy to address selectively subradiant modes ($k_x>k_0$) in a 2D lattice with $k_0d<\pi$ could be to set the array close to a dielectric or a metallic substrate and exploit the evanescent field generated in the total internal reflection or in surface plasmonic wave. Similar approaches have been investigated in Ref.~\citep{Stehle2014,Araujo2024}.
An alternative is to drive a localized section of the array, which will excite many modes at the same time: At later time, only the subradiant modes will survive with a substantial population~\citep{Piovella2025}. However, very few photons are emitted through this protocol (subradiant modes get weakly populated only), making this configuration challenging experimentally.

Finally, while the present work addresses the single-excitation regime, the technique proposed can in principle be extented to states with multiple excitations. More specifically, the possibility to use two-level atoms to create and manipulate multiple-photon states
calls for the extension of single-photon results to the many-body regime~\cite{Walther2022,Zhang2022,Pedersen2023}. In particular, combining the long lifetimes of subradiant modes with the photon-photon interactions achieved through dipolar or Rydberg interactions could allow for the creation of tunable atomic platforms for photon manipulation.

%



\vspace{6pt}

\

\appendix
\section[\appendixname~\thesection]{Approximate solution $\Gamma(k_x,0)$}\label{Appendix:A}
We can obtain an approximate solution of $\Gamma(k_x,k_y)$ of Eq.~(\ref{Gamma:2D:1}) for a lattice which is finite, yet assuming $N_x,N_y$ large. To this end, we change the integration variable from $v_x$ and $v_y$ to $v'_x=v_xdN_x$ and $v'_y=v_ydN_y$, which leads to:
\begin{eqnarray}
\Gamma(k_x,k_y)&=&\frac{3\Gamma_0}{\pi (k_0d)^2}
\sum_{m_x, m_y}\int dv'_x\int dv'_y \frac{\mathrm{sinc}^2(v'_x)\mathrm{sinc}^2(v'_y)}{\sqrt{1-C^2(v'_x,v'_y)}}\nonumber\\
&\times&
\left\{1-\left[\hat d_xC_x(v'_x)+\hat d_yC_y(v'_y)+\hat d_z\sqrt{1-C^2(v'_x,v'_y)}\right]^2\right\}
\label{Gamma:2D:2}
\end{eqnarray}
where $C_\alpha(v'_\alpha)=(k_\alpha-2\pi m_x/d_\alpha-2v'_\alpha/dN_\alpha)/k_0$ and $C^2=C_x^2(v'_x)+C^2_y(v'_y)$. In Eq.~(\ref{Gamma:2D:2}) the dependence on $N_x$ and $N_y$ is in $C_x(v'_x)$ and $C_y(v'_y)$, respectively; furthermore, the $\mathrm{sinc}^2(v'_{x,y})$ is appreciable only for $|v'_{x,y}|< \pi$. We are looking for an approximate solution for the subradiant region of the spectrum under the condition that in the integral of Eq.~(\ref{Gamma:2D:2}) $C^2(v'_x,v'_y)<1$. Let us consider the spectrum along the $k_x$ axis ($k_y=0$), for dipoles perpendicular to the 2D array ($\mathbf{\hat d}=(0,0,1)$). Taking $g_x=g_y=0$, we obtain, up to terms $O(1/N^2_{x,y})$:
\begin{equation}
C^2(v'_x,v'_y)\approx \frac{1}{k_0^2}\left[k_x^2-\frac{4k_x}{dN_x}v'_x\right].
\end{equation}
Inserted in Eq.~(\ref{Gamma:2D:2}), we obtain the rate
\begin{eqnarray}
\Gamma(k_x,0)&=&\frac{3\Gamma_0}{2k_0d}\sqrt{\frac{N_x}{k_xd}}
\int_{v_0}^{+\infty} dv\frac{\mathrm{sinc}^2(v)}{\sqrt{v-v_0}}
\left[\frac{k_x^2}{k_0^2}-\frac{4k_xv}{k_0^2dN_x}\right],
\label{Gamma:2D:3}
\end{eqnarray}
where $v_0=(dN_x/4k_x)(k_x^2-k_0^2)$. In the subradiant region, when $k_0<|k_x|<\pi/d$, 
using $\mathrm{sinc}^2(v)\approx 1/(1+v^2)$ the integrals in Eq.~(\ref{Gamma:2D:3}) can be solved exactly as
\begin{eqnarray}
\Gamma(k_x,0)&=&\frac{3\pi\Gamma_0}{2(k_0d)^3}\left\{(k_xd)^{3/2}\sqrt{N_x}
\frac{\sin[(1/2)\arctan(1/v_0)]}{4(1+v_0^2)^{1/4}}
-4\sqrt{\frac{k_xd}{N_x}}
\sqrt{\frac{v_0+\sqrt{1+v_0^2}}{2(1+v_0^2)}}\right\}.\nonumber\\
\label{Gamma:2D:Appendix:A}
\end{eqnarray}

\section[\appendixname~\thesubsection]{Approximate solution $\Gamma(k_x,0,0)$}\label{Appendix:B}
An approximate solution of $\Gamma(k_x,k_y,k_z)$ from Eq.~(\ref{Gamma:3D:3}) can be obtained for a finite cubic lattice assuming $N_x, N_y, N_z$ large.  In the subradiance region of the spectrum, we consider the spectrum along the direction $k_x$ axis ($k_y=k_z=0$) and for dipoles oriented along the $z$-axis ($\mathbf{\hat d}=(0,0,1)$). Taking $\mathbf{g}=0$, Eq.~(\ref{Gamma:3D:3}) takes the form
\begin{eqnarray}
\Gamma(k_x,0,0)&=&\frac{3\Gamma_0}{\pi k_0^2}N_xN_yN_z
\int dv_x\int dv_y\mathrm{sinc}^2(v_xdN_x)\mathrm{sinc}^2(v_ydN_y)\nonumber\\
&\times &\frac{C^2(v_x,v_y)}{\sqrt{1-C^2(v_x,v_y)}}
\mathrm{sinc}^2\left[\left(k_0\sqrt{1-C^2(v_x,v_y)}\right)dN_z/2\right].
\label{App:3D:1}
\end{eqnarray}
By changing the integration variables as $v_xdN_x\rightarrow v_x$ and $v_ydN_y\rightarrow v_y$, we obtain up to terms $O(1/N^2_{x,y})$
\begin{eqnarray}
\Gamma(k_x,0,0)&=&\frac{3\Gamma_0 N_z}{(k_0d)^2}
\int dv_x\mathrm{sinc}^2(v_x)\nonumber\\
&\times &\frac{V^2(v_x)}{k_0\sqrt{k_0^2-V^2(v_x)}}
\mathrm{sinc}^2\left[\left(\sqrt{k_0^2-V^2(v_x)}\right)dN_z/2\right],
\label{App:3D:2}
\end{eqnarray}
where $V^2(v_x)=k_x^2-4k_xv_x/dN_x$. 
Introducing $\eta=(dN_x/2)(k_x-k_0)$, we can write
\[
k_0^2-V^2(v_x)=\frac{4k_0}{dN_x}(v_x-\eta)-4\frac{\eta^2}{d^2N_x^2}+8\frac{v_x\eta}{d^2N_x^2}\approx \frac{4k_0}{dN_x}(v_x-\eta)
\]
if we neglect the terms proportional to $1/N_x^2$. Then we obtain for the rate
\begin{eqnarray}
\Gamma(k_x,0,0)&=&\frac{3\Gamma_0 N_z}{2(k_0d)^2}
\int_\eta^\infty\frac{\mathrm{sinc}^2(v_x)}{\sqrt{(v_x-\eta)/k_0dN_x}}\left[1-\frac{4(v_x-\eta)}{k_0dN_x}\right]\nonumber\\
&\times &\mathrm{sinc}^2\left[\left(\sqrt{k_0d(v_x-\eta)N_z^2/N_x}\right)\right]dv_x.
\label{App:3D:3}
\end{eqnarray}
Changing the integration variable into $t=\sqrt{k_0d(v_x-\eta)N_z^2/N_x}$ we obtain
\begin{eqnarray}
\Gamma(k_x,0,0)&=&\frac{3\Gamma_0 N_x}{(k_0d)^2}
\int_0^\infty\mathrm{sinc}^2(t)\mathrm{sinc}^2\left[\eta+\frac{N_xt^2}{k_0dN_z^2}\right]\left[1-\frac{4t^2}{(k_0d)^3N_z^2}\right]dt.
\label{App:3D:4}
\end{eqnarray}
If we also neglect the terms $(N_xt^2/k_0dN_z^2)$ and $4t^2/[(k_0d)^3 N_z^2]$ (large size limit), we obtain
\begin{eqnarray}
\Gamma(k_x,0,0)&=&\frac{3\pi\Gamma_0 N_x}{2(k_0d)^2}\mathrm{sinc}^2(\eta)=\frac{3\pi\Gamma N_x}{2(k_0d)^2}\mathrm{sinc}^2[dN_x(k_x-k_0)/2)].
\label{App:3D:5}
\end{eqnarray}

\end{document}